\documentclass[a4paper,11pt]{article}
\pdfoutput=1 

\usepackage{jcappub} 

\usepackage[T1]{fontenc} 

\title{Directionality preservation of nuclear recoils in an emulsion detector for directional dark matter search}

\author[a,b,c,d,1]{A. Alexandrov,\note{Corresponding author.}}
\author[a,b]{G. De Lellis,}
\author[a,b]{A. Di Crescenzo,}
\author[a,b]{A. Golovatiuk}
\author[a]{and V. Tioukov}

\affiliation[a]{I.N.F.N. sezione di Napoli, I-80126 Napoli, Italy}
\affiliation[b]{Università degli Studi di Napoli Federico II, I-80126 Napoli, Italy}
\affiliation[c]{National University of Science and Technology MISIS, RUS-119049 Moscow, Russia}
\affiliation[d]{Lebedev Physical Institute of the Russian Academy of Sciences, RUS-119991 Moscow, Russia}

\emailAdd{andrey.alexandrov@na.infn.it}
\emailAdd{giovanni.delellis@unina.it}
\emailAdd{antonia.dicrescenzo@unina.it}
\emailAdd{artem.golovatiuk@cern.ch}
\emailAdd{valeri.tioukov@na.infn.it}

\abstract{Nuclear emulsion is a well-known detector type proposed also for the directional detection of dark matter. In this paper, we study one of the most important properties of direction-sensitive detectors: the preservation by nuclear recoils of the direction of impinging dark matter particles. For nuclear emulsion detectors, it is the first detailed study where a realistic nuclear recoil energy distribution with all possible recoil atom types is exploited. Moreover, for the first time we study the granularity effect on the emulsion detector directional performance. As well as we compare nuclear emulsion with other directional detectors: in terms of direction preservation nuclear emulsion outperforms the other detectors for WIMP masses above 100 GeV/c$^2$.}

\begin{document}
\maketitle
\flushbottom

\section{Introduction}
\label{sec:intro}

Directional detection of dark matter (DM) is the strategy pursued by both new and some existing experiments in the design of next-generation detectors~\cite{Mayet,UFN_review}. Based on the expected anisotropy of the signal from the DM, this strategy offers a unique opportunity to identify interactions of so-called Weakly Interacting Massive Particles (WIMP) reliably even in the presence of unrecoverable background events. The main objectives of a direction-sensitive detector are both to measure the released energy and to reconstruct the direction of motion of the recoil nucleus after a WIMP scattering. Therefore, the direction-preserving property of the detector becomes extremely important as it is directly connected with its performance.

In order to reconstruct nuclear recoil tracks, the spatial resolution and the granularity of the detector should be smaller than  the track length, which, in turn, depends on the transferred energy and density of the target material. For example, in solids the mean track length of a recoil nucleus with an energy of few tens keV is of the order of few hundreds nm. At present, such spatial resolution is achievable only with nuclear emulsions used by the NEWSdm experiment~\cite{NEWS_LOI}. Therefore, most directional-detection experiments make use of a low-pressure ($\sim$ 0.1 bar) gas (LPG) target, where the tracks of recoil nuclei have a millimeter extent~\cite{Battat}. The use of LPGs significantly relaxes the spatial resolution requirements of the detector, albeit at the price of a much lower mass and, hence, the sensitivity per unit volume.

The sensitivity study of nuclear emulsions for WIMP detection can be found in ref.~\cite{Antonia}, but the property of WIMP direction preservation was not studied there, neither was it compared with other detector types.
The comparison of the MIMAC LPG detector with nuclear emulsions and ZnWO$_4$ scintillator crystals was reported in ref.~\cite{Couturier}, where the simulation is not sufficiently detailed and is misleading. Their analysis considered only one recoiling atom species per detector type for all WIMP masses. Moreover, if the choice of recoiling atoms made in ref.~\cite{Couturier} can be justified for light WIMPs, when the energy transfer to lighter atoms is more efficient, this is not the case for heavier WIMPs where the contribution from heavy atoms in emulsion becomes rapidly dominant and must be taken into account. Therefore, as it will be shown later in this paper, the analysis carried out in ref.~\cite{Couturier} significantly underestimates the directionality preservation property of nuclear emulsion, especially for heavy WIMPs.

With this paper we fix that flaw by generating the realistic energy spectrum for all recoiling atom types for each WIMP mass considered. Combining all possible recoiling atom types with realistic energies, we provide the correct evaluation of nuclear emulsion performance as a directional-sensitive DM detector. We focus our study to WIMP masses in the range from 10 to 1000 GeV/c$^2$, which is the most popular mass range according to the concept of WIMP.

\subsection{Nuclear emulsions}

The nuclear emulsion \cite{Emul_GDL} is composed of tiny silver bromide (AgBr) crystals immersed in a gelatine binder. The crystals act as sensors that are activated by the ionization loss of a passing-through charged particle. The activated state of the crystals is preserved until the emulsion film is chemically developed. Thus, a particle track is recorded, first as a sequence of activated crystals, which later, after the development, becomes a sequence of silver nanoparticles, called grains. These grains have the form of randomly oriented filaments with typical dimensions equal to several tens of nanometers, depending on the emulsion type.

Two new emulsion types with nanometric crystals \cite{Asada} were designed on purpose, for directional DM search.  The first type, called the Nano-Imaging Tracker (NIT), has crystal size of about 44 nm with the average distance between crystals equal to 71 nm. The second emulsion type is called the Ultra-NIT (UNIT) and has crystal size and average distance equal to 25 nm and 40 nm, respectively. The average distance between crystals defines the granularity of the emulsion detector and, hence, the minimal track length with measurable direction.

\subsection{Emulsion readout}

The readout of nuclear emulsions is typically done by means of optical microscopes. For large emulsion detector experiments that could contain tons of nuclear emulsion, for example OPERA \cite{OPERA_det,OPERA_res}, special robotized microscopes are built \cite{LASSO,NGSS} and novel scanning techniques are designed \cite{LASSO_CM,LASSO_IM} for automatization and acceleration of the emulsion readout.

Due to the so-called diffraction limit in optics, tracks shorter than 200 nm are not accessible with conventional optical microscopes. Therefore, a special super-resolution imaging technique and a dedicated robotized optical microscope are designed to readout from NIT and UNIT emulsions \cite{Umemoto,LASSO_SR}. The effective resolution of the existing super-resolution microscope is about 80 nm, comparable to the granularity of the NIT emulsion. For the 3-dimensional reconstruction of recoil tracks the NEWSdm collaboration has designed a 3D super-resolution microscope \cite{3DMic}. The next generation of the super-resolution microscope is now under design and will have the resolution comparable to the granularity of the UNIT emulsion.

\section{WIMP simulation}

\begin{figure}[tbp]
\centering 
\includegraphics[width=.9\textwidth]{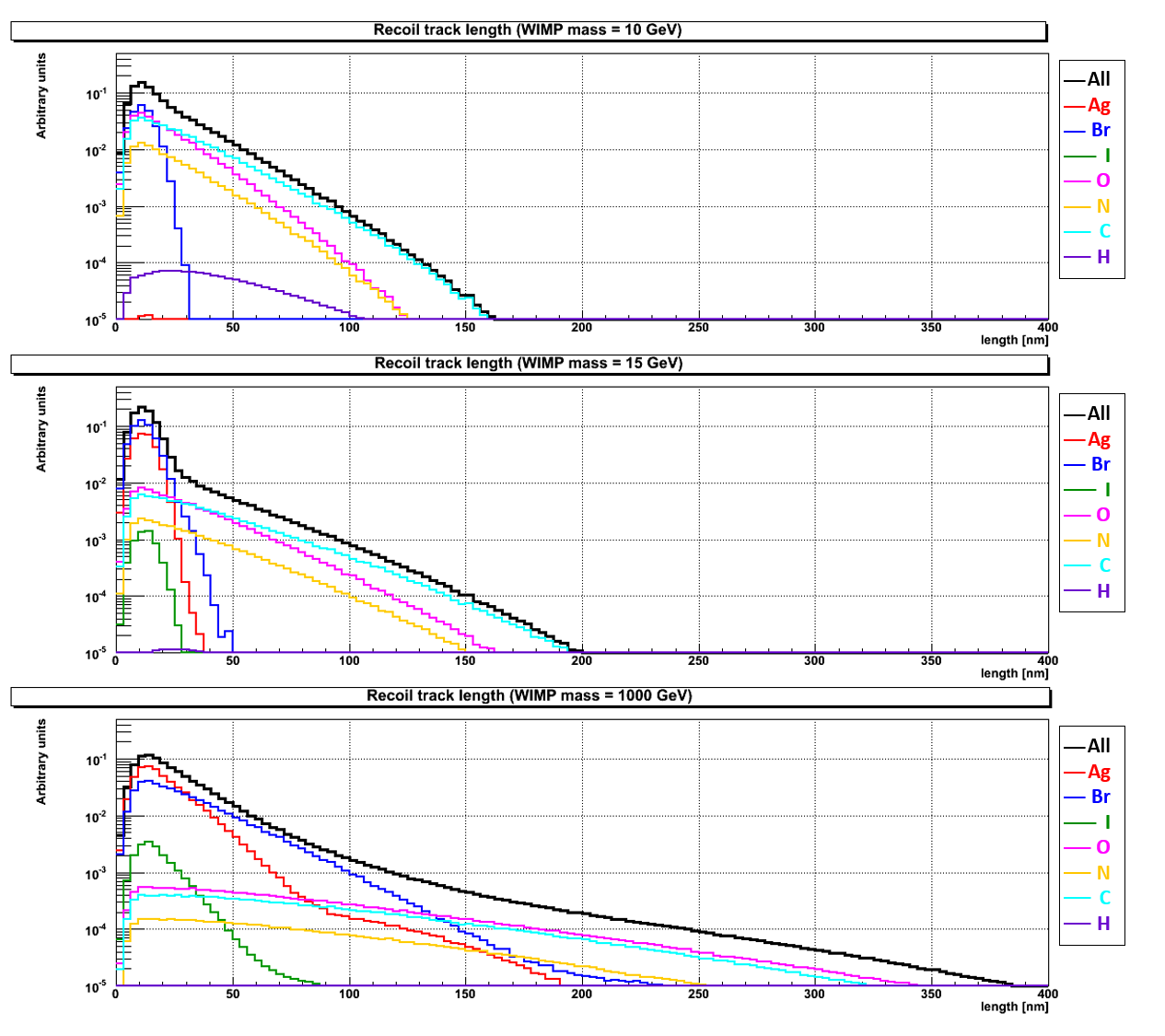}
\caption{\label{fig:1} Individual contributions from different recoiling atoms in emulsion to track length distributions for WIMP masses equal to 10 (top), 15 (middle) and 1000 (bottom) GeV/c$^2$. The plots are normalized to the expected event rate for the corresponding WIMP mass.}
\end{figure}

\begin{figure}[tbp]
\centering 
\includegraphics[width=.9\textwidth]{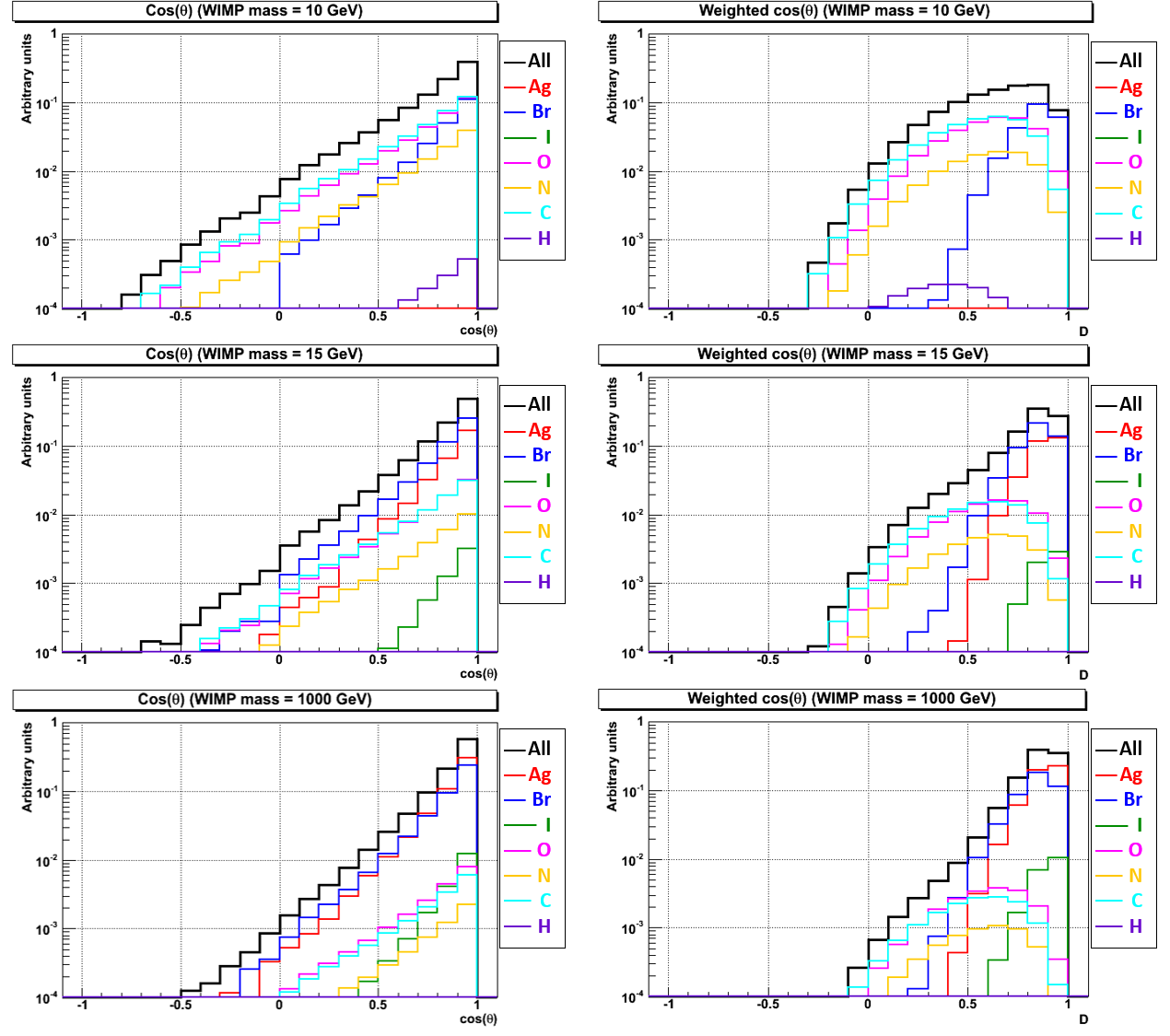}
\caption{\label{fig:2} Individual contributions from different recoiling atoms in emulsion to $\cos{\theta}$ (left column) and $weighted$-$\cos{\theta}$ (right column) distributions in case of WIMP masses equal to 10 (top row), 15 (center row) and 1000 (bottom row) GeV/c$^2$.The plots are normalized to the expected event rate for the corresponding WIMP mass.}
\end{figure}

\subsection{Recoil event rate}
In our study we derived the event rate for WIMP recoils following the approach similar to refs.~\cite{Smith,Alenazi,Gondolo} with more explicit derivation of angular spectrum for recoiled ions.
Following the notations of ref.~\cite{Smith} the analytical expression for the spectrum can be written as:
\begin{equation}
\label{eq:R_etheta}    \frac{d^2 R}{dE\ d\cos\theta} = \frac{R_0}{2N_{esc}\ E_0r} \left[\exp\left(\frac{(v_n-v_E\cos\theta)^2}{v_0^2}\right) - \exp\left(-\frac{v^2_{esc}}{v^2_0}\right) \right] \times F^2(qr_n)
\end{equation}
where $v_0, v_E$ and $v_{esc}$ are, respectively, WIMP velocity dispersion, Earth's velocity relative to the Halo and Galactic escape velocity, $E$ and $\theta$ are energy and angle of the recoiled ion, respectively, and $F(qr_n)$ is the nuclear form-factor (defined below in eq.~\eqref{form-fact}).
\begin{align*}
    v_n &= v_0 \sqrt{E/E_0r}, \ q = \sqrt{2M_T E}\\
    E_0 &= M_W v_0^2/2, \ r = \frac{4M_W M_T}{(M_W+M_T)^2}\\
    N_{esc} &= \text{erf} \left[\frac{v_{esc}}{v_0}\right] + \frac{2}{\sqrt{\pi}} \frac{v_{esc}}{v_0} \exp\left[-\frac{v_{esc}^2}{v_0^2} \right]
\end{align*}
where $M_W$ and $M_T$ are WIMP and target-ion masses, respectively, $v_n$ is the nuclei velocity after the recoil and $E_0$ corresponds to mean kinetic energy of the WIMP.\\
We are using the parameters of the Standard Halo Model and Earth velocity from refs.~\cite{Undagoitia,McCabe}:
\begin{equation*}
    v_0 = 220\text{ km s}^{-1};\ v_E = 232\text{ km s}^{-1}; \ v_{esc} = 544 \text{ m s}^{-1};\ \rho_0 = 0.3\text{ GeV }c^{-2}\text{ cm}^{-3}
\end{equation*}
Together with numerical expressions from ref.~\cite{Smith} and updated Dark Matter parameters:
\begin{gather*}
    R_0 = \frac{2}{\sqrt{\pi}} \frac{N_A}{A}\frac{\rho_0}{M_W}\sigma_0v_0 = \frac{361}{M_W M_T} \left(\frac{\sigma_0}{10^{-36}\text{cm}^2}\right) \left(\frac{\rho_0}{0.3\text{ GeV}c^{-2}\text{cm}^{-3}}\right) \left(\frac{v_0}{220\text{ km s}^{-1}}\right) \text{kg}^{-1}\text{d}^{-1} \\
    E_0r = 2 v_0^2 \frac{M_W^2 M_T}{(M_W+M_T)^2} = \left(\frac{M_W}{100\text{ GeV}c^{-2}}\right) \left(\frac{v_0}{220\text{ km s}^{-1}}\right)^2 \frac{4M_W M_T}{(M_W+M_T)^2} \times 26.9 \text{ keV}
\end{gather*}
where $A$ is the recoiled ion atomic number, $N_A$ is Avogadro number and $\rho_0$ is local Dark Matter density.\\
The cross-section with the nucleus $\sigma_0$ and with single-nucleon $\sigma_n$ cross-section for spin-independent interactions are related by the following formula:
\begin{equation*}
    \sigma_0 = \sigma_{n} A^2\frac{\mu_{_{WT}}^2}{\mu_{_{Wn}}^2}
\end{equation*}
where $\mu_{_{WT}}$ and $\mu_{_{Wn}}$ are the WIMP-target and WIMP-nucleon reduced masses. \\
The Helm nuclear form factor approximation suggested in ref.~\cite{Smith}:
\begin{equation}
\label{form-fact}    F(qr_n) = 3\frac{\sin (qr_n)-qr_n \cos(qr_n)}{(qr_n)^2} e^{-(qs)^2/2}
\end{equation}

$r^2_n = c^2 + \frac{7}{3} \pi^2 a^2-5s^2;\ c = 1.23 A^{1/3} -0.6 \text{ fm};\ a = 0.52 \text{ fm};\ s = 0.9 \text{ fm.}$ 

We are using eq.~\eqref{eq:R_etheta} to calculate spectra for ions in different detectors for the WIMP masses between 10 GeV/c$^2$ and 1000 GeV/c$^2$.

\subsection{Detector parameters}
For the comparison between detector types, similar datasets for the same WIMP masses were generated for emulsion, LPG and crystal detectors. All possible recoil atom types and the realistic energy spectrum corresponding to each detector type were considered. \\
Detector compositions and densities for LPG and crystal detectors are taken from ref.~\cite{Couturier} while for the emulsion they are taken from ref.~\cite{Asada}, namely:
\begin{itemize}
    \item \textbf{LPG detector}: $70\% \text{CF}_4$ + $28\%\text{CHF}_3$ + $2\%\text{C}_4\text{H}_{10}$ gas mixture with the density of 1.72$\cdot 10^{-4}$g/cm$^3$. Atomic fraction: H(0.0927), C(0.2046), F(0.7027).
    \item \textbf{Crystal scintillators}: ZnWO$_4$ with the density of 7.87 g/cm$^3$. Atomic fraction: Zn(0.167), O(0.666), W(0.167).
    \item \textbf{NIT and UNIT emulsion}: the density is 3.44 g/cm$^3$ and atomic fraction: H(0.41), C(0.214), N(0.049), O(0.117), Br(0.101), Ag(0.105), I(0.004).
\end{itemize}

\begin{table}[tbp]
\centering
 \begin{tabular}{|c | c || c | c |} 
 \hline
 Ion & $E_{max}$ (keV) & Ion & $E_{max}$ (keV) \\ 
 \hline
 H & 12.4 & Zn & 717.9\\
 \hline
 C & 145.2 & Br & 855.4 \\
 \hline
 N & 168.7 & Ag & 1100.8 \\
 \hline
 O & 192 & I & 1254.3 \\
 \hline
 F & 224.5 & W & 1656.2 \\
 \hline
\end{tabular}
\caption{\label{tab:Emax}The highest energies for ions used in SRIM simulations}
\end{table}

\subsection{SRIM simulation}
We perform the simulation in two steps:
\begin{itemize}
    \item Generating ions of fixed initial energy and angle in the corresponding detector for bins in the energy-angle grid.
    \item Using the energy and angular spectrum from eq.~\eqref{eq:R_etheta} to weigh the contribution of the ions in specific bins to the resulting distribution.
\end{itemize}

The bins for energy-angles grid are defined for each ion as energies from 0 keV to $E_{max}$ - maximum recoil energy produced by 1000 GeV/c$^2$ WIMP, for which eq.~\eqref{eq:R_etheta} becomes zero, and angles from 0 to $\pi/2$. The highest energies for each recoil atom type are shown in table~\ref{tab:Emax}. The weigh for each energy-angle bin is calculated as a ratio of the differential recoil rate from eq.~\eqref{eq:R_etheta} times bin size $(\Delta E\Delta\theta)$ the full WIMP rate of a specific WIMP mass in the corresponding detector.

\subsection{Simulation cross-check with data}

The recent comparison of the simulation with the experimental data can be found in ref.~\cite{LASSO_SR}, where the endpoint of the simulated recoil track length distribution for C ions with energy of 60 keV showed a good agreement with the measured one.

\begin{figure}[tbp]
\centering 
\includegraphics[width=.7\textwidth]{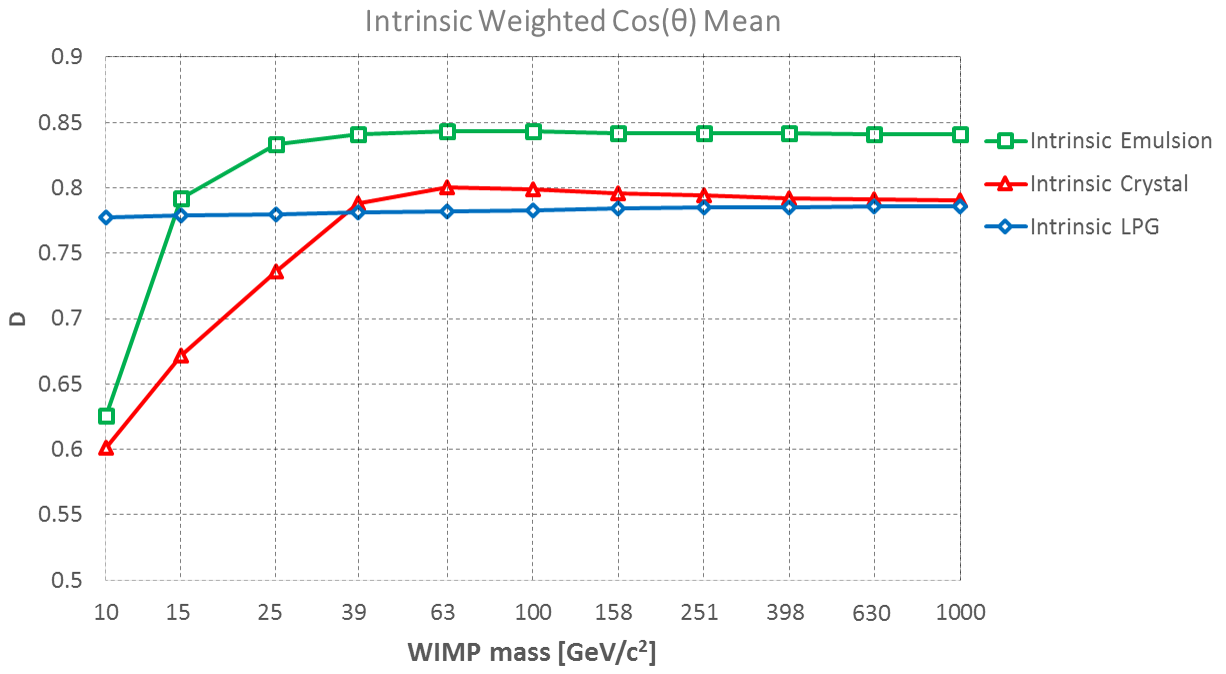}
\caption{\label{fig:3} $Weighted$-$\cos{\theta}$ mean values as a function of the WIMP mass for LPG (blue diamonds), crystal (red triangles) and emulsion (green squares) detectors.}
\end{figure}

\begin{figure}[tbp]
\centering 
\includegraphics[width=.9\textwidth]{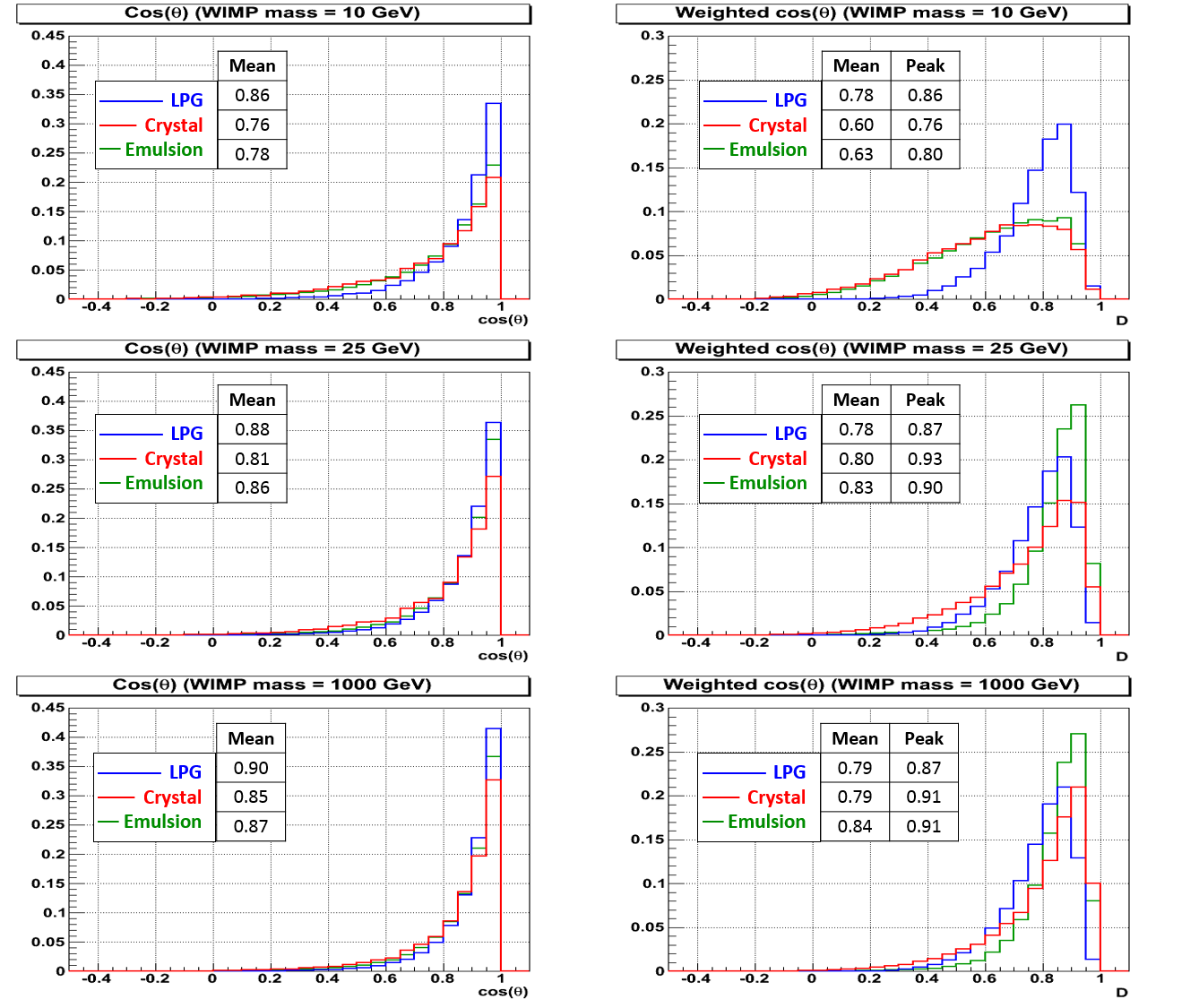}
\caption{\label{fig:4} Intrinsic distributions for $\cos{\theta}$ (left column) and $weighted$-$\cos{\theta}$ (right column) of LPG (blue line), crystal (red line) and emulsion (green line) detectors for WIMP masses equal to 10 (top row), 25 (middle row) and 1000 (bottom row) GeV/c$^2$. The plots are normalized to the expected event rate for the corresponding WIMP mass and detector.}
\end{figure}

\section{Results}

We have simulated several datasets for WIMP masses ranging from 10 to 1000 GeV/c$^2$. For each dataset the recoil angle $\theta$, its cosine and the energy-weighted cosine distribution $D$, introduced in ref.~\cite{Couturier}, were calculated. Similarly to ref.~\cite{Couturier}, $\theta$ is defined as the angle between the line connecting the initial and the stopping points of the track and its initial direction. The energy-weighted $\cos{\theta}$ distribution is referred to as the "Figure~of~Merit" in ref.~\cite{Couturier}. However, for the current analysis we prefer to call it the "$weighted$-$\cos{\theta}$ distribution $D$". For calculation we use the same formula defined in ref.~\cite{Couturier} and for convenience we report it also here:

\begin{equation*}
D=\frac{\sum_{i=0}^{N_{collisions}} \Delta{E_i} \cos{\theta_i}}{\sum_{i=0}^{N_{collisions}} \Delta{E_i}} = \frac{\langle \Delta{E} \cos{\theta} \rangle_{track}}{\langle \Delta{E} \rangle_{track}}
\end{equation*}
where $\theta_i$ is the angle between the direction of the initial recoil and the line joining two subsequent interaction points $i$ and $i + 1$ (i.e. the direction after the $i$-th collision); $\Delta{E_i}$ is the energy deposited by the recoil between $i$ and $i + 1$; $N_{collisions}$ is total number of interactions/collisions.

\subsection{Individual atomic contributions in emulsion}

Individual contribution from each recoiling atom type in emulsion to the track length, $\cos{\theta}$ and $weighted$-$\cos{\theta}$ distributions for WIMP masses ranging from 10 to 1000 GeV/c$^2$ are shown in figure~\ref{fig:1} and figure~\ref{fig:2}. The plots are normalized to the expected event rate for the corresponding WIMP mass.  

In right panels of figure~\ref{fig:2}, if we consider separately distributions for light and heavy atoms, we see that their shape does not change with WIMP mass. Heavier atoms always perform better at any WIMP mass. However, their contribution in emulsion at masses below 15 GeV/c$^2$ is small due to a relatively low event rate. As the event rate for heavy atoms increases with the WIMP mass, their relative contribution to the overall detector performance also increases, thus improving the overall performance. Their contribution becomes dominant at about 25 GeV/c$^2$, thus saturating the detector performance. This behaviour is clearly visible in figure~\ref{fig:3}, and it is valid for all detector types. Therefore, an emulsion detector can operate in three different regimes, motivating the choice of WIMP masses in figure~\ref{fig:1} and figure~\ref{fig:2}: 
\begin{enumerate}
    \item light-atoms dominated mode for WIMP masses below 10 GeV/c$^2$;
    \item mixed mode for WIMP masses between 10 GeV/c$^2$ and 25 GeV/c$^2$;
    \item heavy-atoms dominated mode for WIMP masses above 25 GeV/c$^2$.
\end{enumerate}
As it will be shown later, the granularity effect shifts the limits of applicability for each operation regime towards higher WIMP masses.

\subsection{Intrinsic detector performance}

In this paragraph we introduce the intrinsic detector performance, i.e. the ideal performance achievable on the basis of its nuclear composition, without accounting for the granularity of the detector. The dependency of mean values of the intrinsic $weighted$-$\cos{\theta}$ on WIMP masses ranging from 10 to 1000 GeV/c$^2$  is plotted in figure~\ref{fig:3}. The comparison of intrinsic $weighted$-$\cos{\theta}$ and angular distributions for WIMP masses equal 10, 25 and 1000 GeV/c$^2$ is shown in figure~\ref{fig:4}. As expected, the LPG detector has the best performance at WIMP mass around 10 GeV/c$^2$. However, at higher WIMP masses both emulsion and crystal detectors rapidly fill the gap and supersede it, as the relative contribution from heavy atoms increases. Above 25 GeV/c$^2$ the intrinsic mean $weighted$-$\cos{\theta}$ of the emulsion detector is about 5-7\% higher than that of both the LPG and crystal detectors, making the emulsion detector an excellent choice for the directional WIMP detection in that mass range.

\begin{figure}[tbp]
\centering 
\includegraphics[width=.9\textwidth]{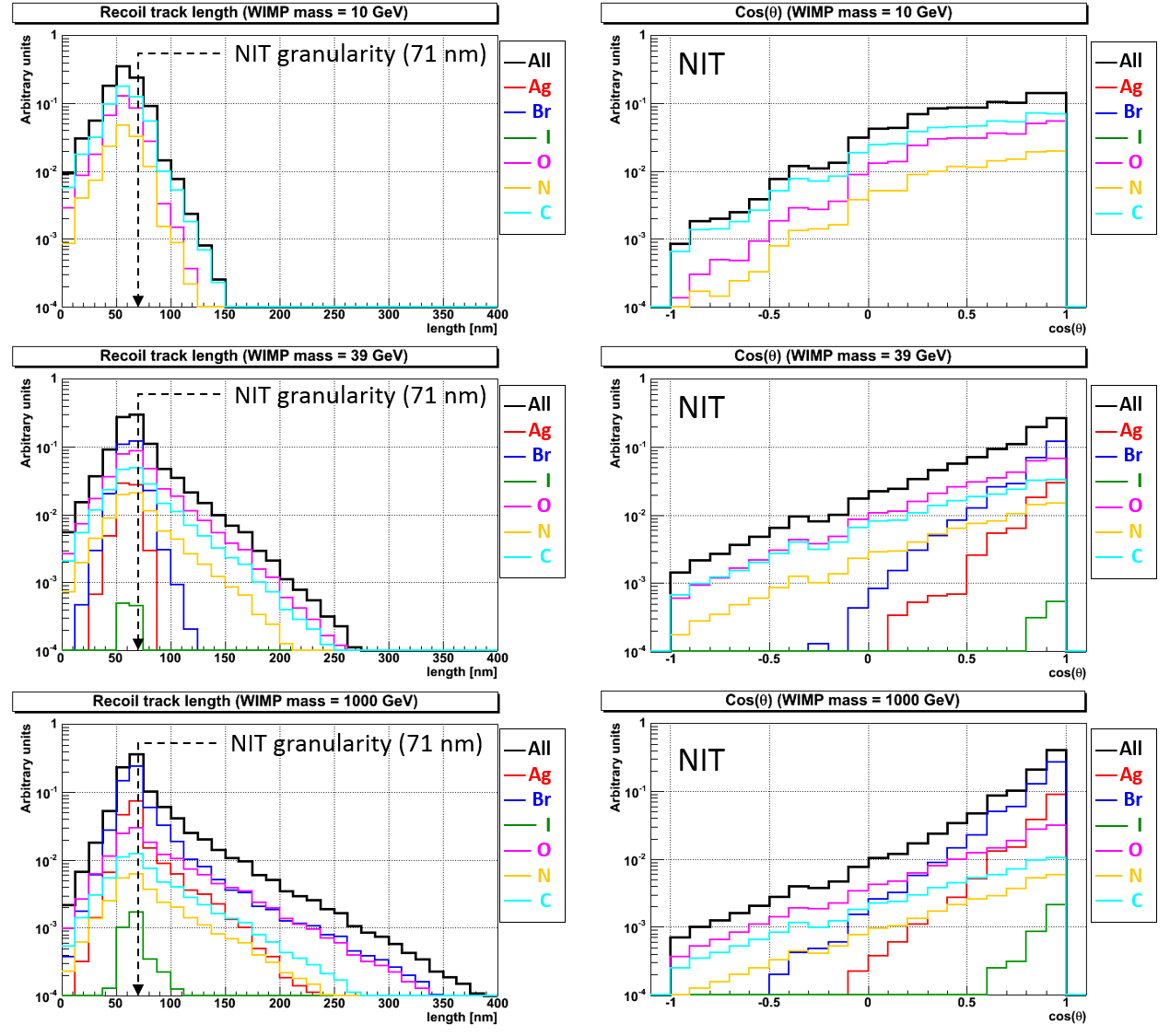}
\caption{\label{fig:5} Recoil track length (left column) and $\cos{\theta}$ (right column) distributions in NIT emulsion for WIMP masses equal to 10 GeV/c$^2$ (top row), 39 GeV/c$^2$ (middle row) and 1000 GeV/c$^2$ (bottom row). The plots are normalized to the event rate for the corresponding WIMP mass.}
\end{figure}

\begin{figure}[tbp]
\centering 
\includegraphics[width=0.9\textwidth]{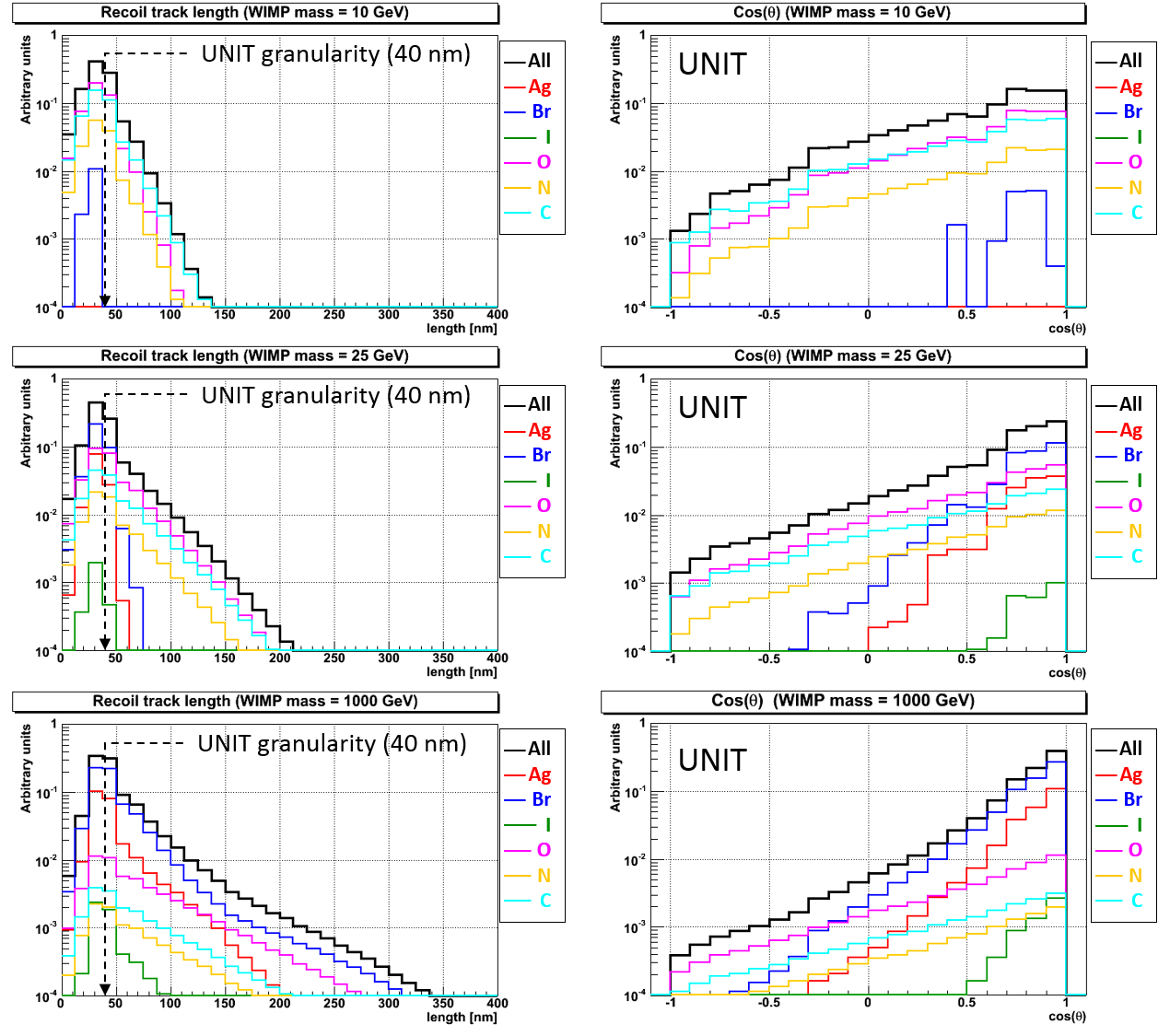}
\caption{\label{fig:6} Recoil track length (left column) and $\cos{\theta}$ (right column) distributions in UNIT emulsion for WIMP masses equal to 10 GeV/c$^2$ (top row), 25 GeV/c$^2$ (middle row) and 1000 GeV/c$^2$ (bottom row). The plots are normalized to the  event rate for the corresponding WIMP mass.}
\end{figure}

\begin{figure}[tbp]
\centering 
\includegraphics[width=0.75\textwidth]{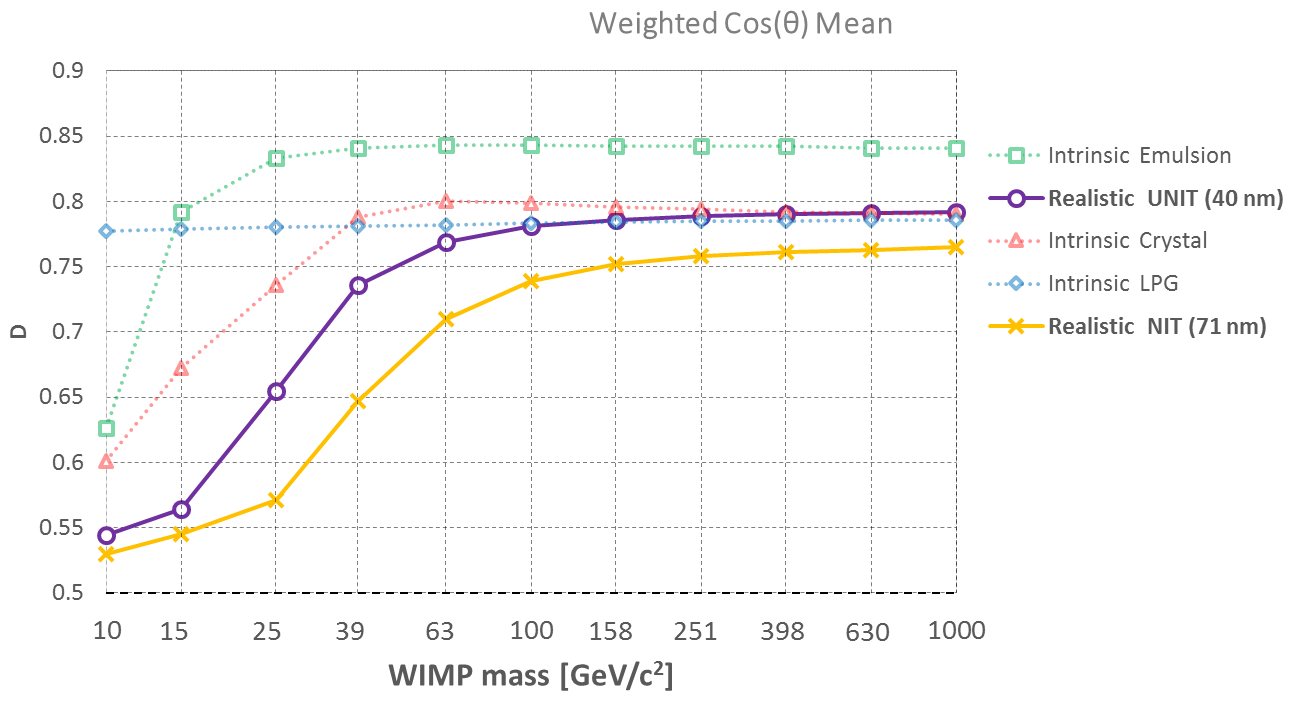}
\caption{\label{fig:7} Realistic distribution of mean values of $\cos{\theta}$ for NIT (orange crosses) and UNIT (violet circles) emulsions. Also shown for comparison intrinsic distributions of mean values of $weighted$-$\cos{\theta}$ for emulsion (green squares), crystal (red triangles) and LPG (blue diamonds) detectors.}
\end{figure}

\subsection{Granularity effect on detector performance}

In order to be reconstructed in emulsion the recoiling ions must encounter at least two AgBr crystal while travelling in an emulsion detector and release inside them the amount of energy sufficient for activation. 
We approximate the nuclear emulsion as a "crystal field": a 3D volume randomly filled with crystals with the average center-to-center distance equal to 71$\pm$5 nm and 40$\pm$3 nm for NIT and UNIT, respectively. 
Each crystal is approximated with a sphere of diameter equal to 44$\pm$7 nm and 25$\pm$4 nm for NIT and UNIT, respectively.  
Then, we conduct recoiling ions through this crystal field keeping track of crystals crossed and calculating the energy deposited inside them. 
The crystal activation threshold is set equal to 2 keV.

The result of this procedure is shown in figure~\ref{fig:5} and  figure~\ref{fig:6} for NIT and UNIT emulsions, respectively. Recoil track length distributions become peaked slightly below the corresponding granularity value. This is due to that the realistic track length distribution is a convolution the exponentially falling intrinsic one and the gaussian-shaped center-to-center distance distribution. Moreover, both NIT and UNIT emulsions become not sensitive to proton recoils as the latter are not capable of releasing enough energy inside AgBr crystals to activate them.
Another consequence of the exponentially falling track length distribution is that the most of recoiling ions activate only two crystals and, therefore, the $weighted$-$\cos{\theta}$ distribution becomes equal to the $\cos{\theta}$ one.

The granularity effect reduces the performance of both NIT and UNIT emulsion detectors due to a suppression of the contribution from heavy Ag and Br atoms, whose ranges are, on average, shorter than those of lighter C, N and O atoms.
As a result, the contribution from light atoms keeps dominating until heavier WIMP masses and the limits of all operation regimes are correspondingly shifted towards higher WIMP masses: this justifies the choice of 39 GeV/c$^2$ and 25 GeV/c$^2$ for the mixed mode plots in figure~\ref{fig:5} and  figure~\ref{fig:6} for NIT and UNIT detectors, respectively.
As it can be seen from figure~\ref{fig:7}, the directionality preservation performance of the NIT emulsion is less than 5\% lower than the intrinsic one of the LPG for WIMP masses above 100 GeV/c$^2$, while the UNIT slightly supersedes the LPG in the same mass range. Application of emulsions with granularities finer than 40 nm will further improve the directionality preservation property of emulsion by pushing it closer to the intrinsic limit.

\section{Discussion}

In the presented study we have analyzed the directionality preservation properties for different detector types: LPG, ZnWO$_4$ scintillator crystals and nuclear emulsions. The developed SRIM-based simulation considers all possible recoil atom types and takes into account the realistic recoil energy spectrum for WIMPs with masses in the range from 10 to 1000 GeV/c$^2$. Our simulation shows that the LPG performance is almost flat in the considered WIMP mass range. Indeed, the gas mixture contains only light atoms of similar masses and, therefore, their relative contribution does not change much with the WIMP mass. On the contrary, nuclear emulsion contains a large amount of heavy Ag and Br atoms. At low WIMP masses the major contribution into the detected signal comes from light C, N and O atoms. Naturally, being the lightest in the mixture they suffer from stronger scattering. However, the contribution from Ag and Br rapidly becomes dominant and notably improves the emulsion performance. Disregarding this fact in ref.~\cite{Couturier} resulted in significant underestimation of emulsion capabilities especially for heavy WIMPs. With the presented detailed analysis we demonstrate that in terms of intrinsic directionality preservation, if granularity effects are not considered, nuclear emulsion outperforms other detector types for directional DM search in the WIMP mass range above 15 GeV/c$^2$.

The random crystal field model enabled consideration of the granularity effect and the estimation of realistic performances of the existing emulsion detectors.
Both NIT and UNIT performances show some drop at low WIMP masses due to the suppression of the contribution from heavy ions. Nevertheless, they show excellent performance at higher WIMP masses, comparable to the ideal intrinsic performances of both LPG and ZnWO$_4$ detectors. 
Taking into account that the realistic performances of the latter detectors are expected to be worse, UNIT emulsions are complementary to other directional detection technologies since they perform better for masses heavier than 100 GeV/c$^2$.
Production of emulsions with granularity finer than 40 nm is possible and can further improve the directional preservation performance of emulsion detectors pushing it closer to the intrinsic limit. Another important advantage of a nuclear emulsion detector is that, being solid-state, it can be easily scaled to larger masses in a compact volume without major safety implications and does not require complex multichannel instrumentation for long and stable operation.


\acknowledgments

This work is supported by a Marie Sklodowska-Curie Innovative Training Network Fellowship of the European Commissions Horizon 2020 Programme under contract number 765710 INSIGHTS.



\end{document}